\documentclass{article}

\usepackage{arxiv}
\usepackage{amsmath}

\usepackage[utf8]{inputenc} 
\usepackage[T1]{fontenc}    
\usepackage{hyperref}       
\usepackage{url}            
\usepackage{booktabs}       
\usepackage{amsfonts}       
\usepackage{nicefrac}       
\usepackage{microtype}      
\usepackage{lipsum}         
\usepackage[utf8]{inputenc}
\usepackage[T1]{fontenc}
\usepackage{graphicx,booktabs}
\usepackage{appendix}
\usepackage{colortbl}
\usepackage{setspace}
\usepackage{chngcntr}
\usepackage{authblk}
\usepackage{titlesec}
\usepackage{afterpage}
\usepackage{url}
\usepackage{cleveref}       
\usepackage{dcolumn}
\usepackage{verbatim}
\usepackage[export]{adjustbox}
\usepackage{chngpage}
\usepackage{floatrow}
\newfloatcommand{capbtabbox}{table}[][\FBwidth]

\usepackage[tableposition=top]{caption}
\usepackage{subcaption}

\usepackage{fancyhdr}
\usepackage{natbib}

\urldef{\viewsurl}\url{viewsforecasting.org}
\usepackage{longtable}
\usepackage{float}
\usepackage{import}
\usepackage{booktabs}
\usepackage{comment}
\usepackage{lscape}
\usepackage{rotating}
\usepackage[hashEnumerators,smartEllipses]{markdown}

\usepackage{listings}
\usepackage{bbm}

\usepackage{tabulary}

\begin{document}

\title{\large The 2023/24 VIEWS Prediction challenge: \\ Predicting the number of fatalities in armed conflict, with uncertainty\footnote{The research was funded by the German Ministry of Foreign Affairs, Uppsala University, the European Research Council project ERC-AdG 101055175 (ANTICIPATE), Riksbankens Jubileumsfond (Societies at Risk), and the Center for Advanced Studies, Oslo.}
}

\author[1, 2]{\footnotesize{{Håvard Hegre}}}
\author[1, 2]{\footnotesize{{Paola Vesco}}}
\author[2, 3]{\footnotesize{{Michael Colaresi}}}
\author[1]{\footnotesize{{Jonas Vestby}}}
\author[1]{\footnotesize{{Alexa Timlick}}}
\author[1]{\footnotesize{{Noorain Syed Kazmi}}}
\author[4]{\footnotesize{{Friederike Becker}}} 
\author[5]{\footnotesize{{Marco Binetti}}}
\author[4]{\footnotesize{{Tobias Bodentien}}}
\author[5]{\footnotesize{{Tobias Bohne}}}
\author[6]{\footnotesize{{Patrick T. Brandt}}}
\author[7]{\footnotesize{{Thomas Chadefaux}}}
\author[4]{\footnotesize{{Simon Drauz}}}
\author[8]{\footnotesize{{Christoph Dworschak}}}
\author[9]{\footnotesize{{Vito D'Orazio}}}
\author[10]{\footnotesize{{Cornelius Fritz}}} 
\author[7]{\footnotesize{{Hannah Frank}}}
\author[11,1]{\footnotesize{{Kristian Skrede Gleditsch}}} 
\author[5]{\footnotesize{{Sonja Häffner}}}
\author[ ]{\footnotesize{{Martin Hofer}}}
\author[12]{\footnotesize{{Finn L. Klebe}}}
\author[13]{\footnotesize{{Luca Macis}}}
\author[14]{\footnotesize{{Alexandra Malaga}}} %
\author[15]{\footnotesize{{Marius Mehrl}}} 
\author[12]{\footnotesize{{Nils W. Metternich}}} 
\author[5]{{\footnotesize{Daniel Mittermaier}}}
\author[16]{\footnotesize{{David Muchlinski}}} 
\author[14,17]{\footnotesize{{Hannes Mueller}}} 
\author[5]{\footnotesize{{Christian Oswald}}}
\author[13]{\footnotesize{{Paola Pisano}}} 
\author[2]{\footnotesize{{David Randahl}}} 
\author[18]{\footnotesize{{Christopher Rauh}}} 
\author[4]{\footnotesize{{Lotta Rüter}}} 
\author[7]{\footnotesize{{Thomas Schincariol}} }
\author[19]{\footnotesize{{Benjamin Seimon}}} 
\author[13]{\footnotesize{{Elena Siletti}}}
\author[13]{\footnotesize{{Marco Tagliapietra}}} 
\author[16]{\footnotesize{{Chandler Thornhill}}} 
\author[20]{\footnotesize{{Johan Vegelius}}} 
\author[5]{\footnotesize{{Julian Walterskirchen}}}

\affil[1]{\footnotesize{Peace Research Institute Oslo (PRIO)}}
\affil[2]{\footnotesize{Department of Peace and Conflict Research, Uppsala University}}
\affil[3]{\footnotesize{University of Pittsburgh}}
\affil[4]{\footnotesize{Institute of Statistics (STAT), Karlsruhe Institute of Technology (KIT)}}
\affil[5]{\footnotesize{Center for Crisis Early Warning, University of the Bundeswehr Munich}}
\affil[6]{\footnotesize{School of Economic, Political, and Policy Sciences, University of Texas, Dallas}}
\affil[7]{\footnotesize{Trinity College Dublin}}
\affil[8]{\footnotesize{University of York}}
\affil[9]{\footnotesize{West Virginia University}}
\affil[10]{\footnotesize{Pennsylvania State University}}
\affil[11]{\footnotesize{University of Essex}}
\affil[12]{\footnotesize{University College London}}
\affil[13]{\footnotesize{Department of Economics and Statistics "Cognetti de Martiis", University of Turin}}
\affil[14]{\footnotesize{Institute for Economic Analysis, Barcelona}}
\affil[15]{\footnotesize{University of Leeds}}
\affil[16]{\footnotesize{Georgia Tech}}
\affil[17]{\footnotesize{Barcelona School of Economics}}
\affil[18]{\footnotesize{University of Cambridge}}
\affil[19]{\footnotesize{Fundació Economia Analítica}}
\affil[20]{\footnotesize{Department of Medical Sciences, Uppsala University}}

\date{\footnotesize{\today}\\
}
\maketitle

\begin{centering}
\vspace{10mm}
\includegraphics[width=.15\linewidth]{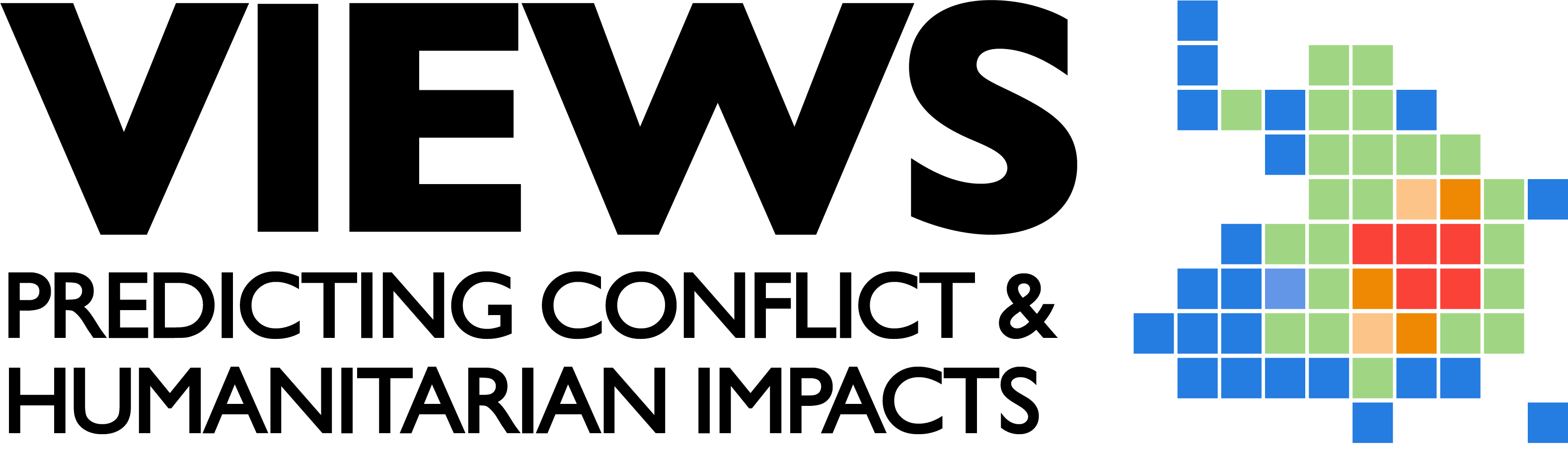} \hspace{2mm}
\includegraphics[width=.15\linewidth]{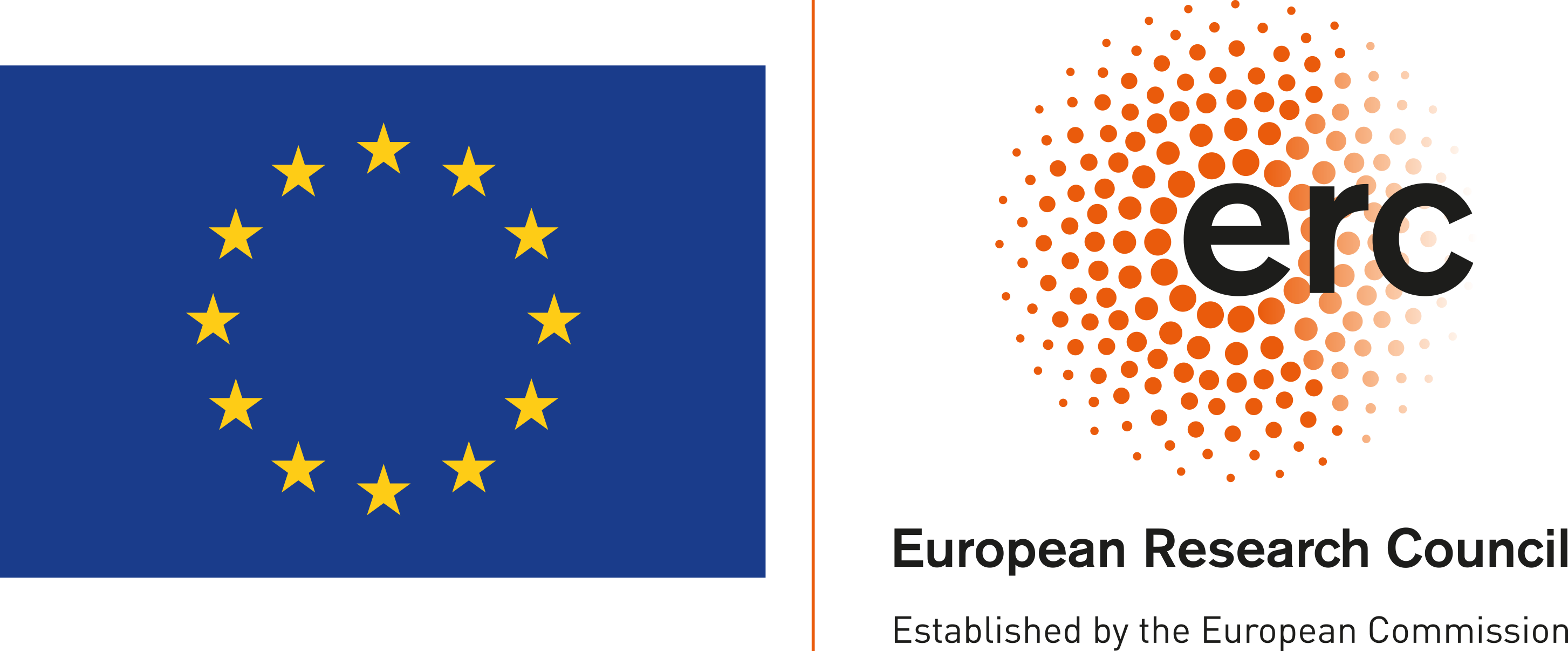}
\includegraphics[width=.15\linewidth]{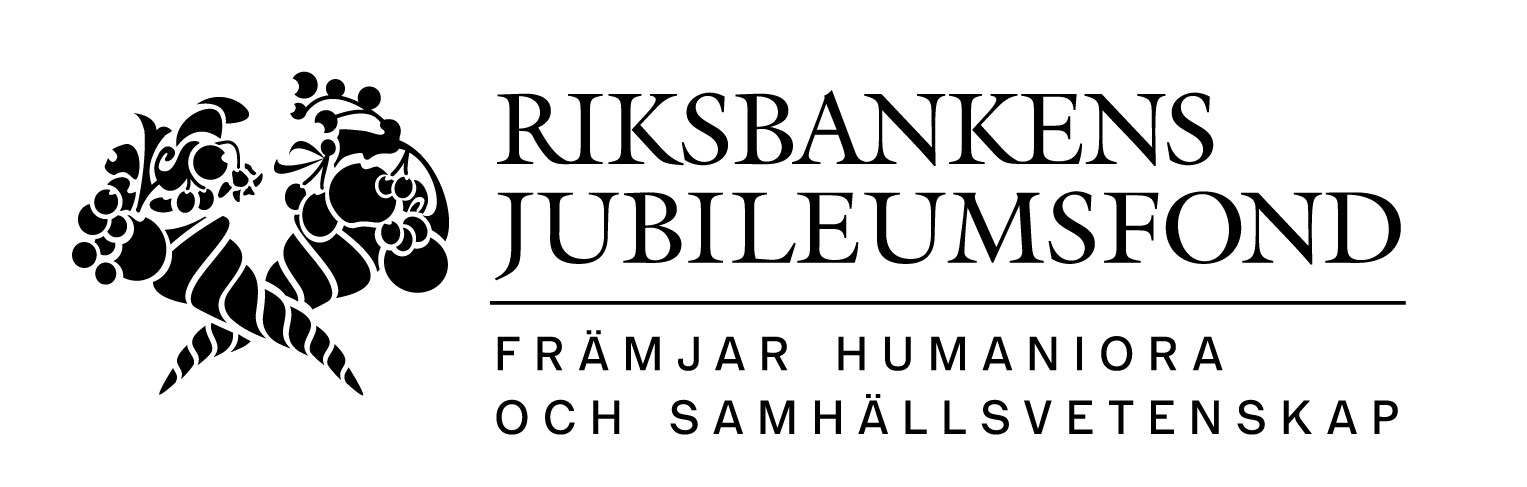}  
\includegraphics[width=.15\linewidth]{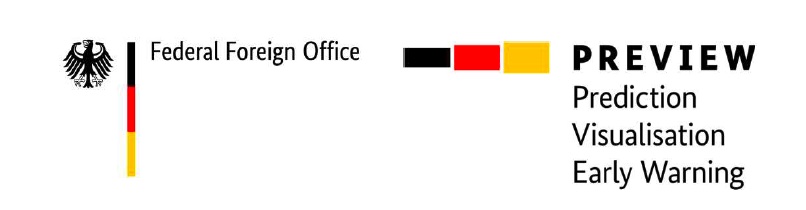}  
\includegraphics[width=.15\linewidth]{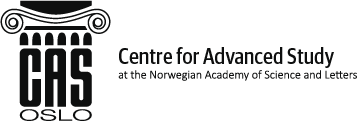}
\includegraphics[width=.15\linewidth]{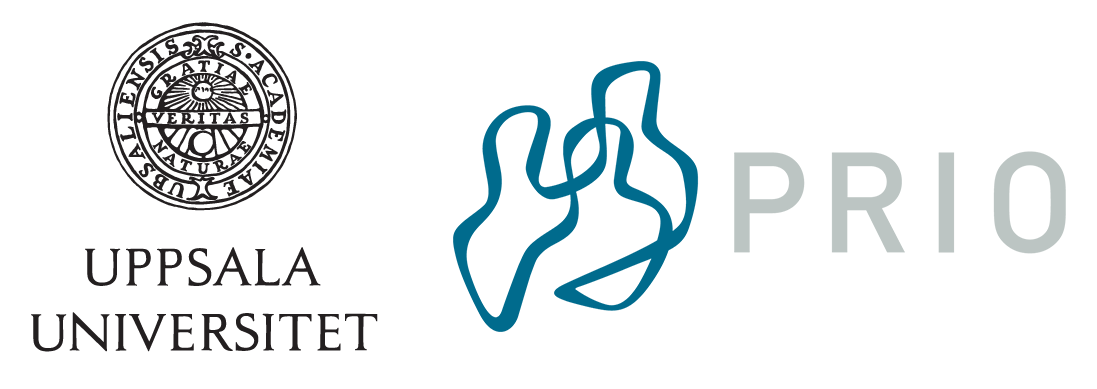} 
\end{centering}

\begin{abstract}
\noindent\footnotesize{ This draft article outlines a prediction challenge where the target is to forecast the number of fatalities in armed conflicts, in the form of the UCDP `best' estimates, aggregated to the VIEWS units of analysis. It presents the format of the contributions, the evaluation metric, and the procedures, and a brief summary of the contributions. The article serves a function analogous to a pre-analysis plan: a statement of the forecasting models made publicly available before the true future prediction window commences. More information on the challenge, and all data referred to in this document, can be found at \url{https://viewsforecasting.org/research/prediction-challenge-2023/}.}

\end{abstract}

\section{Motivation}
Since the \citet{UNWB2017} report calling for `early warning-early action' procedures, armed conflict forecasts have been in increasing demand within IGOs such as the UN or within governments. Several such organizations are developing early-warning systems, or including forecasts from systems like the Violence Early Warning System \citep[ViEWS --][]{Hegre2019JPR, Hegre2021JPR} in their `dashboards' supporting decision making. To the extent that the forecasts are sufficiently precise, they are used either to stimulate efforts to prevent conflict escalation, or, more realistically, to support efforts to mitigate their consequences. Typically, users are both interested in the most likely outcome (a point prediction), but also in the lower-probability risk that conflicts escalate catastrophically (the tail ends of the probability distribution). Point predictions are arguably not very informative in the situations where early warnings are most useful, namely before large-scale violence erupts in places where there has been limited previous violence. Since it can take long time before severe tension escalates into overt violence, the best point prediction tends to cluster around no violence. Forecasts in the form of probability distributions, on the other hand, can alert to a low but alarming risk of a large-scale conflagration. Users are also interested in knowing how uncertain the forecasts are. As such, predictions of war intensity as probability distributions comes closer to what user groups need than point estimates or simple dichotomous predictions from classification models \citep{Brandt2014IJF}.

Generating forecasts as probability distributions, however, is new to the field of armed conflict forecasting. Most of the existing efforts, in fact, provide forecasts as point predictions without any measures of uncertainty \citep[e.g.][]{Goldstone:2010AJPS, Brandt:2011CMPS, Blair2020JCR, Hegre:2013ISQ, Chiba2017JPR, Mueller2018APSR, Dorff2020JOP, Bazzi2022RES, Hegre2022II, Vesco2022II}. To strengthen the knowledge basis for such modeling, the VIEWS project has invited researcher teams interested in forecasting models in general and in the prediction of armed conflict in particular, to take part in a prediction challenge where all contributors work on a common, well-defined task:

\textit{To predict the number of fatalities in armed conflict, as reported by the Uppsala Conflict Data Program (UCDP), with estimates of the uncertainty of the predictions calculated in the form of samples of forecasted values.}

In this article, we describe the challenge in more detail, present the task, outline the 24 contributions from 13 teams, present the evaluation metrics that they will be scored on, as well as a set of benchmark models that the contributions will be scored against.\footnote{The format of contributions and evaluation metrics were announced in April 2023 \citep[see][ and \url{https://viewsforecasting.org/research/prediction-challenge-2023/}]{Hegre2023PC}.} By rewarding contributions that perform well both in terms of point prediction and uncertainty estimation, the challenge encourages inter-disciplinary efforts to model uncertainty around armed conflict forecasts, increases our understanding of the model characteristics that are most useful to improve probabilistic forecasts, sheds light on the issues involved in evaluating such forecasts, and suggests a set of evaluation metrics that could address these issues.   

This article is supplemented by an interactive visualization tool available at \url{https://predcomp.viewsforecasting.org} to explore all forecasts. In-depth summaries for all models can be found at \url{https://viewsforecasting.org/research/prediction-challenge-2023/}.

\section{The challenge}
The challenge builds on the predecessor VIEWS prediction competition \citep{Hegre2022II,Vesco2022II}, where the task was to predict \textit{change} in the number of conflict fatalities. The previous competition taught us some valuable lessons on the value of forecasting conflict fatalities, while raising some important limitations and challenges. It was clear that complex models based on sophisticated algorithms and leveraging big data are the best individual tool to predict changes in fatalities -- although they tend to be difficult to interpret \citep{Vesco2022II}. Even very sophisticated machine learning models, however, tend to be surprised by the outbreak of conflicts in previously peaceful locations: most (but not all) of the models in fact are beaten, on common forecast evaluation metrics, by a basic `no-change' model that constantly predicts a null change in fatalities. 

\begin{figure}
    \centering
    \includegraphics[width=.49\linewidth]{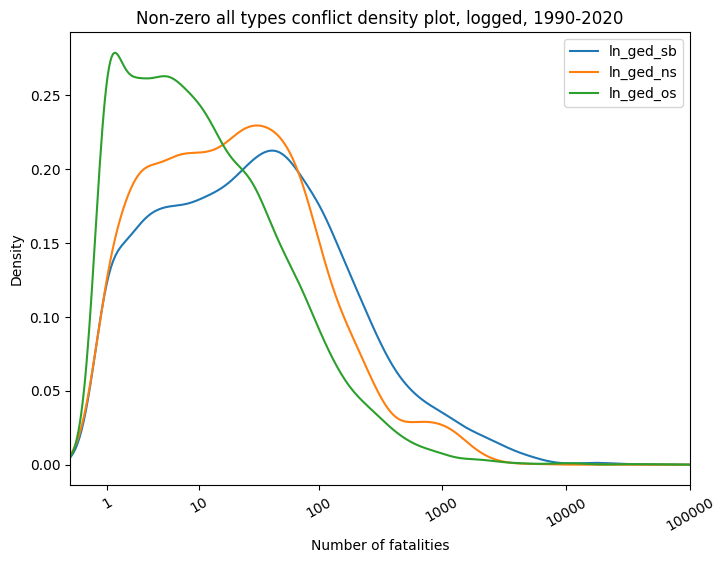}
    \includegraphics[width=.49\linewidth] {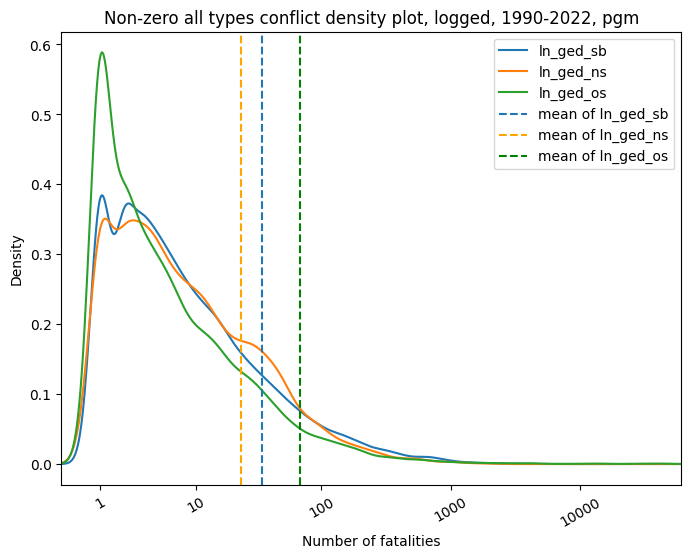} 
    \caption{Distribution of (logged) observed fatalities in December 2022 at our two levels of analysis, \textit{cm} (left) and \textit{pgm} (right), and for different types of violence reported by the UCDP-GED Dataset \citep{Davies2022JPR}: state-based (\textit{``ln\_ged\_sb''}, blue), non-state (\textit{``ln\_ged\_ns''}, orange) and one-sided (\textit{``ln\_ged\_os''}, green).}
    \label{fig:FatalityDistribution}
\end{figure}

These findings suggest that more research is needed to improve our collective ability to forecast conflict outbreaks and (de-)escalation, but also calls for better evaluation metrics to more meaningfully assess each model's accuracy as well as the performance of combined ensembles of models. Point-estimate predictions -- which rely on a single measure of the predicted outcome -- are difficult to evaluate in a way that creates new insights. Point estimates, by definition, do not encode the inherent uncertainty in the distribution of conflict fatalities for a given instance. Moreover, point-estimate predictions make it more difficult to reveal mis-calibration and obscure the relation between the shape and location of the predictive distribution and the underlying data generation process. This divergence might be most pronounced for processes that produce skewed distributions of target observations, such as in the UCDP conflict fatalities we use \citep{Davies2023JPR}. Figure \ref{fig:FatalityDistribution} demonstrates the skewness of the (non-zero) fatality counts at the two VIEWS levels of analysis, the `country-month' (\textit{cm}) and the `PRIO-GRID-month' (\textit{pgm}) levels. Even when plotting the counts on a log axis, the distributions have a distinct skew. To add to this, 87\% of all the observations at the \textit{cm} level are zeros, and as many as 99\% at the \textit{pgm} level.
At the core of this problem is that learning the expected mean of the fatality count distribution --  what evaluation metrics such as the Mean Squared Error (MSE) favors -- may not be of preeminent relevance. 
For an outcome like the one of interest here, with its zero-inflated and right-skewed distribution/probability mass function, the expected value is not sufficiently informative to balance the risks between zero and extremely high values.\footnote{There are infinitely many distributions with the same mean that assign different plausibilities to zero-observations and very large observations. Thus the same expected value (eg a mean of 1.2) can describe a relatively benign situation (eg a 60 percent change of 1 fatality and a 40 percent change of 0 fatalities)} vs a situation that might have a non-ignorable risk of a larger event (eg a 1 percent change of 120 fatalities and a 99 percent change of 0 fatalities).  
If the goal is to produce an insightful characterization of the predictive distribution, which can meaningfully represent both the near-zero and the extreme cases at the tail, then there is a need to move from single-valued point forecasts to probabilistic forecasts that explicitly represent the distribution of plausibility across these (and other values). 
An important lesson of the past competition \citep{Hegre2022II} was therefore the need to go beyond point-estimate predictions, and account for relative certainty and uncertainty across the potential values that fatality observations can take, with evaluation metrics that capture different use cases, or different qualities as we call them below. This is what we seek to explore in this challenge.
\begin{figure}
    \centering
    \includegraphics[width=.49\linewidth]{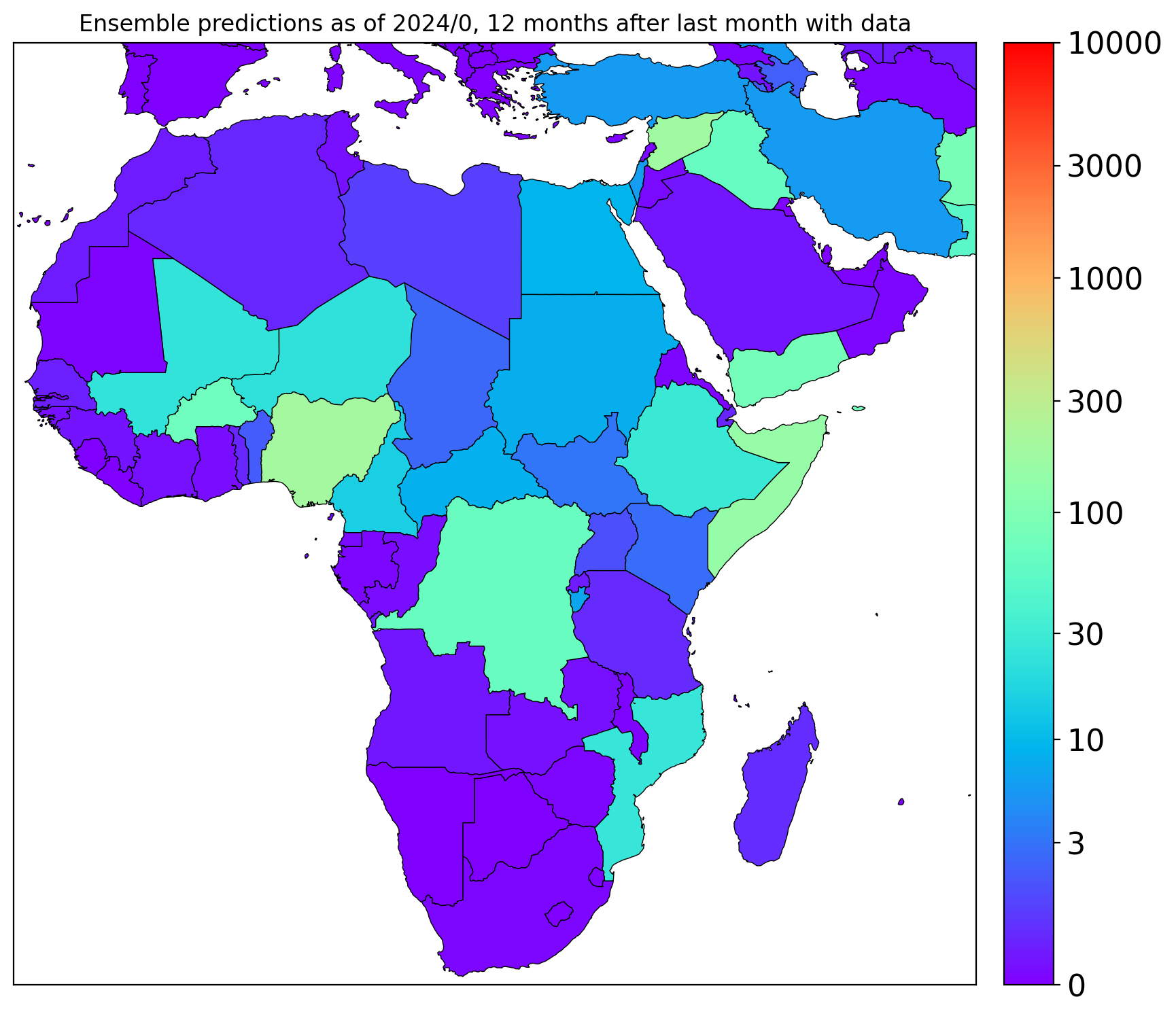}
    \includegraphics[width=.49\linewidth]{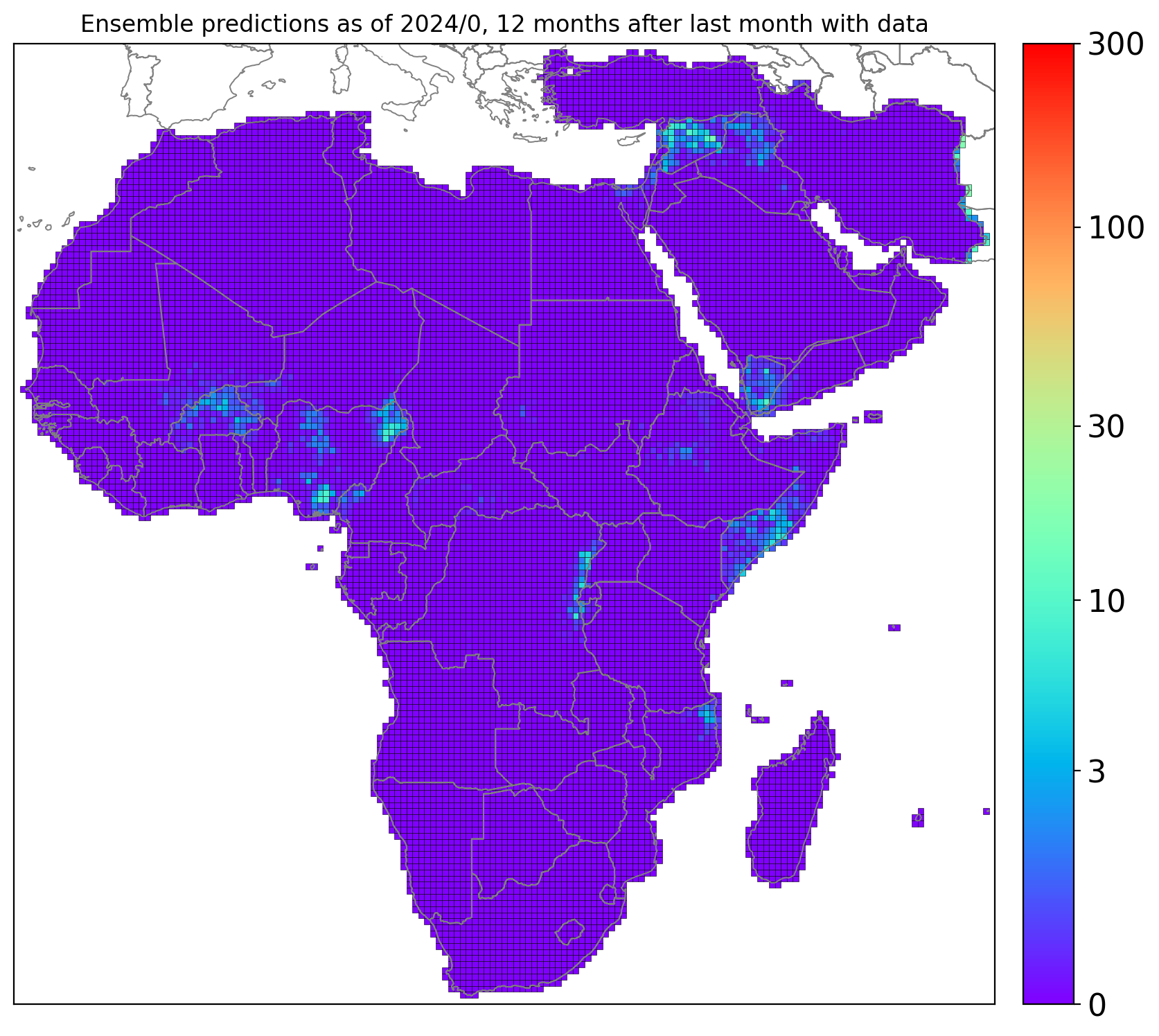}
    \caption{ViEWS ensemble point predictions for January 2024, \textit{cm} and \textit{pgm} level}
    \label{fig:PointPredictions}
\end{figure} \\

\subsection{Prediction target: Predicting the monthly number of fatalities from organized political violence} \label{sec:Target}
The prediction challenge consists of two parts with targets defined at two different geographical units, and contributors can choose to submit contributions at either the country level (\textit{cm}) with global coverage, the sub-national level (PRIO-GRID-month, \textit{pgm}) for Africa and the Middle East\footnote{This geographic scope is dictated by the ViEWS infrastructure which is currently available for Africa and Middle-East only at the PRIO-GRID level. An expansion of the system to global forecasts at the PRIO-GRID level is ongoing and will likely be in place in 2025.}, or both. The temporal resolution at both the \textit{cm} and \textit{pgm} levels is the month. The two parts will have roughly the same structure.

\subsubsection{1+6 prediction windows}
We requested contributions for two sets of prediction windows. The primary goal was to provide predictions for the true future -- the 12 months from July 2024 to June 2025, based on data up to and including April 2024. The predictions were to be provided separately for each of the twelve months in this time window.\footnote{Contributors may decide whether they want to submit identical predictions for each month, or fine-tune predictions separately for each of them.} However, a single year of events at these aggregation levels will generate a limited amount of data for model evaluation, and the evaluation scores will vary considerably over time as the benchmark model evaluation below suggests. To complement the evaluation of the true future forecasts, we also request contributions for each of the calendar years 2018--2023, based on data up to and including October for the preceding year. Seven sets of input data were made available to the participants, one for each of these forecasting windows.
\subsubsection{State-based conflict}
The prediction target is the Uppsala Conflict Data Program (UCDP)'s coding of the number of fatalities in state-based armed conflict, as defined in \citet{Davies2023JPR}. For the test prediction windows up to and including 2023, we have access to the final UCDP data as reported in that article. For the year of 2024 up to April, the UCDP Candidate data will be available \citep{Hegre2020RP}. Contributions were requested to present predictions as a number of draws from the predicted distribution of fatality counts.\footnote{In the \citet{Hegre2022II} competition, the target was specified in log form. There are some compelling reasons for making forecasts in the count form: Adding 1 to $y$ to allow for (log) zeros is an arbitrary choice. Since the vast majority of cases are zero, the choice we apply will make a difference. We believe that evaluating models on the original, non-logged scale most likely, and usefully, rewards models that are willing to inch up away from 0. We may still calculate some auxiliary evaluation metrics based on a log-transformed version.}        
\begin{figure}
    \centering
        \includegraphics[width=.49\linewidth]{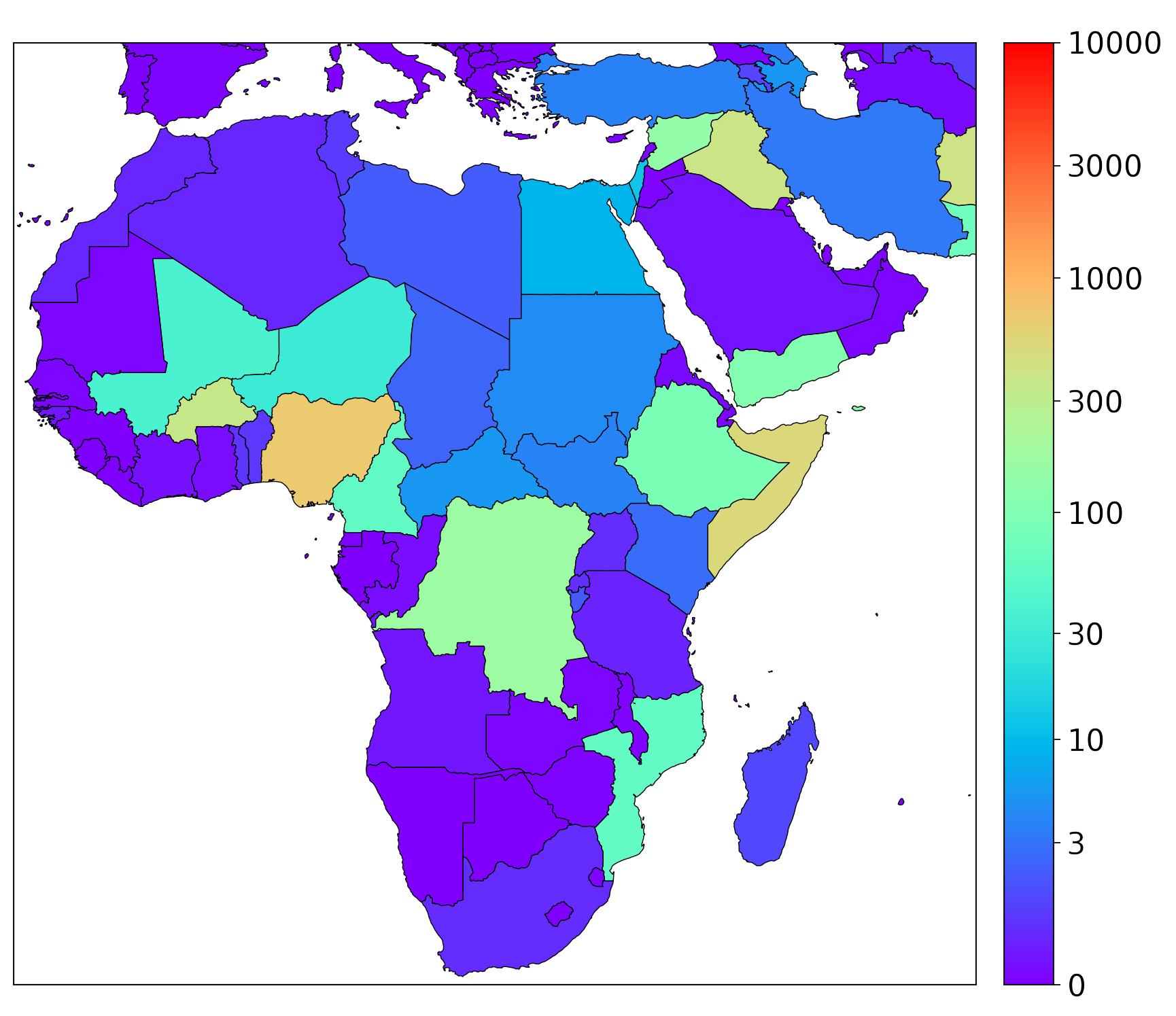}
        \includegraphics[width=.49\linewidth]{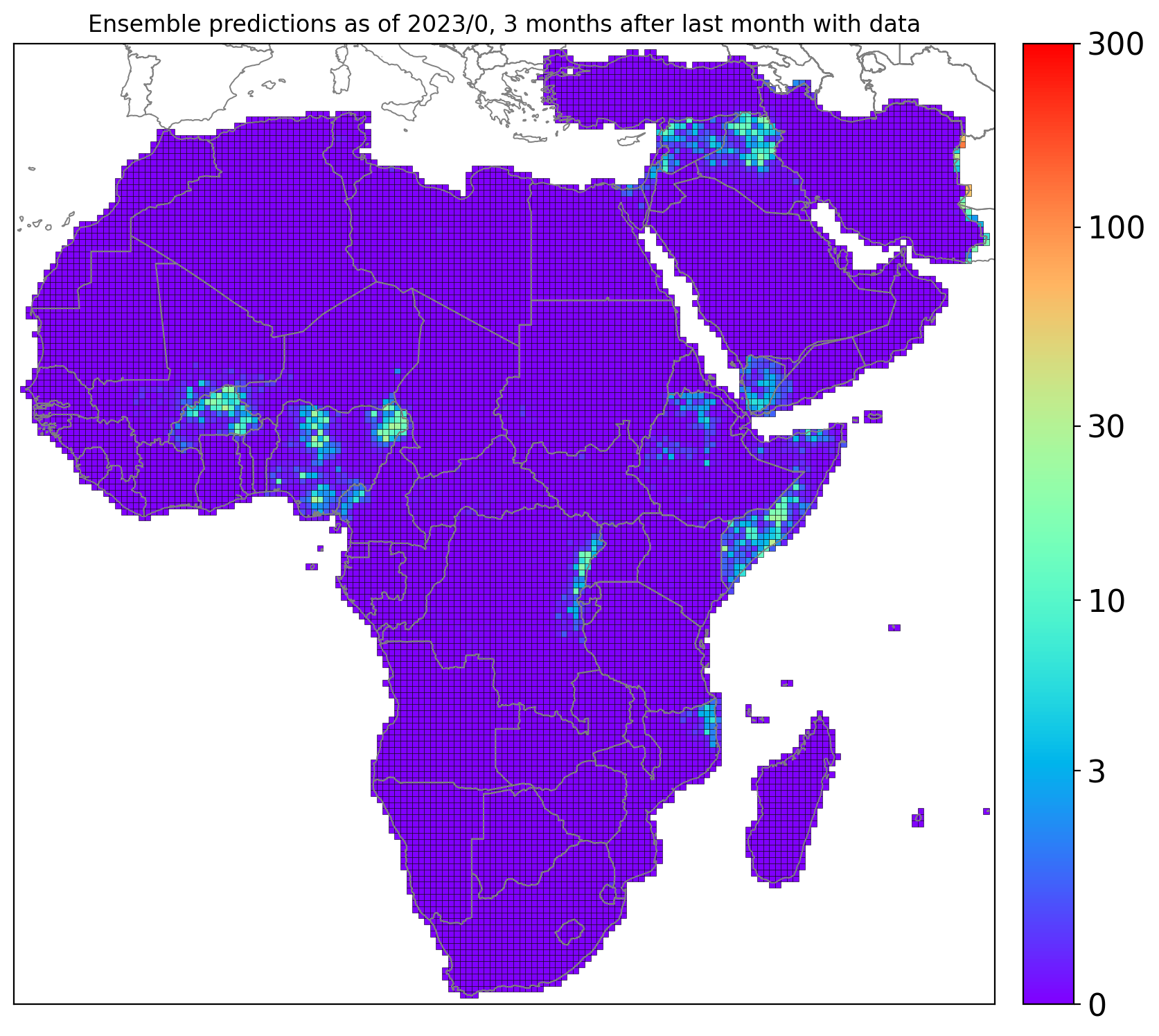} 
    \begin{footnotesize}
    \caption{Predicted fatalities in December 2022 at our two levels of analysis}
    \end{footnotesize}
    \label{fig:Levels}
\end{figure}
\subsubsection{The VIEWS levels of analysis}
The contributors are requested to submit forecasts in either or both of the two VIEWS levels of analysis \citep{Hegre2019JPR, Hegre2021JPR}, as used in the former ViEWS prediction competition \citep{Hegre2022II, Vesco2022II} and depicted in Figure \ref{fig:PointPredictions}:
\paragraph{Country-months}
\citep[][abbreviated \textit{cm} in VIEWS]{Gleditsch:1999II}. The set of countries is defined by the Gleditsch-Ward country code \citep[][with later updates]{Gleditsch:1999II}, and the geographical extent of countries by the latest version of CShapes \citep{Weidmann2010II}. For the country level of analysis VIEWS data are global.
\paragraph{PRIO-GRID months} (\textit{pgm}), which rely on PRIO-GRID \citep[version 2.0;][]{Tollefsen2012JPR}, a standardized  spatial grid structure consisting of quadratic grid cells that jointly cover all areas of the world at a resolution of 0.5 x 0.5 decimal degrees. Near the equator, a side of such a cell is 55 km. For the subnational level of analysis, we currently restrict forecasts to Africa and the Middle East.\footnote{Note that the \textit{cm} and \textit{pgm} definitions are not fully compatible with each other. PRIO-GRID provides a 1:1 cell-to-country correspondence by assigning the grid cell to the country taking up the largest area \citep{Tollefsen2012PG}. When PRIO-GRID cells span two or more countries, all events contained in that PRIO-GRID cell are aggregated, ignoring which country they actually took place in. In the country-month dataset, such events are assigned to the country where the event took place. Moreover, PRIO-GRID cells exist for the entire duration of the dataset, but only those months in which a country has existed in the \citet{Gleditsch:1999II} country list are included in the \textit{cm} datasets.}
\subsection{Timeline and format of contributions} \label{sec:Timeline}
The organizers set the following timeline for the project:

\begin{itemize}
     \item \textbf{Deadline 1}: 1 June 2023: Participants submit abstracts to organizers. We received a total of 20 abstracts, all of which were invited to submit models and forecasts.
    \item \textbf{Deadline 2}: 25 September 2023: Contributors submit their  preliminary forecasts, updated 200-word summaries, and  1500-word summary papers presenting their methods and preliminary forecasts for four test windows (2018--2021). These papers and results were presented in a workshop in Berlin on 10--11 October 2023.
    \item \textbf{Deadline 3}: 23 June 2024: Contributors submit the final predictions for all test windows (2018--2023) and the true future (July 2024 ($s=3 $)--June 2025 ($s=14 $)), as well as updated summaries of their models.
    \item 1 July 2024: Start of prediction window.
    \item 30 June 2025: End of forecasting window.
\end{itemize}
For each test-set observation, a team can provide either (or both) of the following:
\begin{enumerate}
    \item for each observation, indexed by 
    either \[ month\_id, country\_id, draw\] or \[ priogrid\_id, month\_id, draw \]
    supply between 15 and 1000 samples (int32) from their predictive distribution for $Y_{it}$ (where $i$ is either $country\_id$ or $priogrid\_id$ and $t$ is $month\_id$). We (the organizers) take these samples and calculate the necessary functions for each of our evaluation metrics. Contributions in the form of samples are convenient, as they allow the scoring team to evaluate properly the tails of the distributions, and it is also practically easier to fit empirical CDFs/pdfs with samples when the contributed models can potentially generate arbitrary distributions of predictions (i.e. do not follow a specific distribution). See the benchmark models provided at \url{https://viewsforecasting.org/research/prediction-challenge-2023/} for examples.
    \item for each observation $i,t$ supply a point prediction for $Y_{it}$. We generate a set of samples using this point predictions by assuming that the predictive distribution follows a Poisson distribution with the point prediction provided $i,t$ as mean and variance, and evaluate using the same functions as if we had received samples directly from the team. We offer this contribution format to prospective participants that choose to focus on other aspects of the modeling than the shape of the predictive distribution. The Poisson model probably understates the uncertainty of the predictions -- contributors concerned with this were invited to submit predictions as draws from their predictive distribution.
\end{enumerate}
Contributors submitted their predictions as .parquet files indexed by country or PRIO-GRID id, month id, and draw id.\footnote{See Apache Parquet project, \url{https://parquet.apache.org}. There are packages and functions to read and write parquet files in all major open source computing languages commonly used for data science and computational social science work, including python, Julia, and R.} 

Ahead of the kick-off of the challenge, the VIEWS team made the following data available to the contributors, for the \textit{cm} and \textit{pgm} level:
\begin{itemize}
    \item Training data on the predictors for the 1990--2017 period, including about 100 features each at \textit{cm} level and \textit{pgm}. 
    \item Prediction data for the test period, i.e. the data needed to produce forecasts for 1--12 months forward, for all months in the period 2018--2021.
    \item Benchmark models as probability distributions.
\end{itemize}
 In June 2024, contributors got access to updated data up to and including April 2024, necessary to produce predictions for the true future partition.

\section{The contributions}

At the time of writing, the teams have submitted brief summaries outlining their modeling strategies. All of these are briefly outlined below, along with citations for more extensive presentations of the models written by each contributor team. In the fall of 2025, after the end of the true future forecasting window, a separate article will provide a comprehensive evaluation of the performance of the models, based on both the test window forecasts and the 12 months of 2024/25. The short names of models given below are the names identifying them at \url{https://predcomp.viewsforecasting.org}.



\paragraph{\textbf{Drauz\_Becker\_quantile\_forecast	}}	Equally spaced empirical quantiles of each country's most recent historical observations of fatalities, where the number of observations is optimized separately for each combination of country and forecast horizon. The reliance on quantiles rather than random samples guarantees an adequate representation of the predictive distribution, while the non-parametric nature of the approach and the individual adaptation to each location through the tuning parameter ensures its ability to capture specific characteristics of the predictive distribution such as zero-inflation, as well as recent temporal variations. While this baseline-type model is interpretable, transparent, and easy to understand, it is by definition unable to predict any changes in trends
(\cite{drauz_summary_2024}).

\paragraph{\textbf{Bodentien\_Rueter\_NB}} Quantiles from a negative binomial distribution whose parameters are estimated from historical observations of fatalities for each country. The number of observations is optimized for each country separately by minimizing the CRPS in the test set. The resulting model accounts for the overdispersion inherent in the data.  It is simple, straightforward, transparent and easy to interpret. It naturally models the conflict trap characteristic, yet by construction it is unable to predict the outbreak of conflicts or identify trends that have not occurred in the past. The model is compared to more complex approaches, including a hurdle model and a neural network (\cite{bodentien_summary_2024}).
\paragraph{\textbf{CCEW\_tft	}}	Temporal fusion transformer models (TFTs) produce probabilistic predictions through quantile regressions, offering a non-parametric approach to estimate uncertainty. They can incorporate past and future inputs and time-invariant covariates in addition to other exogenous time series to simultaneously produce forecasts for the entire prediction horizon. Furthermore, TFTs  offer interpretability through temporal self-attention decoders, producing variable importance scores and detection of seasonality patterns and extreme events (\cite{walterskirchen_enhancing_2024}).
 \paragraph{\textbf{CCEW\_trees\_global }} A quasi-hurdle ensemble of tree-based models, combining binary classification predictions for the occurrence of fatalities and sample outputs derived from distributional regressors trained on non-zero targets. For each prediction timestep, the modeling pipeline automatically tunes multiple classifiers (Random Forest, XGBoost) and regressors (Distributional Random Forests, Quantile Regression Forests, NGBoost), selecting the best algorithm based on tuning performance. Probabilistic predictions of fatalities are then obtained by interpreting the probability of the classifier as the share of non-zero-predictions in the final samples drawn from the regressor, while the remainder is filled with zeros. \textbf{CCEW\_trees\_global} is the preferred specification, which trains on all available training data. Additional specifications include an ensemble of multiple local models (\textbf{CCEW\_trees\_local}) accounting for regional differences in conflict dynamics through spatial clustering of grid cells derived from geographic distributions of violence and a combined global-local model (\textbf{CCEW\_trees\_global\_local}), where the best combination of global and local hurdle components is selected based on past performance (\cite{mittermaier_forests_2024}).

\paragraph{\textbf{ConflictForecast   }}	A random forest regressor applied on conflict history features as well as 15 topics that are extracted from a large news corpus using a Latent Dirichlet Allocation algorithm. The approach leverages on topics as predictors,  derived from summarizing over 6 million newspaper articles. Country-month uncertainty estimates are obtained by sampling predicted errors based on the predicted probability of conflict and the predicted number of fatalities using a Tweedie distribution. PRIO-GRID uncertainty is obtained through a Quantile Forest Regression (\cite{malaga_predicting_2024}). 
\paragraph{\textbf{Dorazio\_ensemble    }}
Automated machine learning is used to produce point predictions employing conflict history variables constructed from GED state-based conflict events as input features. The dependent variable is a Box-Cox transformation of state-based fatalities, which is back-transformed to the original scale after prediction. Uncertainty estimates are provided by random draws from the Tweedie distribution with mean equals to the point prediction for the PRIO-GRID. The ensemble model uses forecasts from five different outcome transformations, selecting which to use based on a separate forecasting model trained to predict when to use which transformation. An alternative version, the \textbf{Dorazio\_log} model, uses the log transformed outcome ($\lambda=0$ for the Box-Cox transformation) (\cite{dorazio_probabilistic_2024}). 
\paragraph{\textbf{HCD\_dyad	}}	A Random Forest algorithm is trained on clusters of conflict actors obtained through with Dynamic Time Warping and hierarchical clustering. Actor-level clusters capture the temporal and spatial dependencies and can help uncover fighting patterns. The main aim of this approach is to address heterogeneity in grid-level predictions that stem from particular armed organization dyads present in a grid or their spatial proximity. Uncertainty estimates are provided by sampling the outcome distributions from the point predictions and the respective standard errors in each grid, drawing from a truncated random normal distribution (\cite{gleditsch_random_2024}). 
\paragraph{\textbf{Muchlinski\_Thornhill\_zeroinf\_GAM	}} Two-stage modelling approach where the first stage predicts conflict occurrence as a binary variable at the country-month level, and the second stage predicts the number of fatalities conditional on more than zero fatalities being predicted in the first stage. The two stages are combined in a hurdle-like model which is a zero-inflated generalized additive model. In the first stage, the predictions are obtained through a stacked ensemble of machine learning models including a support vector machine, random forests, a neural network, and a penalized logistic regression. In the second stage, the predictions of this ensemble are utilized to overcome the problem of zero-inflation, whereas historical fatality data including three-, six-, and eight-month rolling averages of fatalities along with temporal lags are utilized to forecast the total number of fatalities for each country-month. For the true future forecasts, samples are obtained by simulating 1000 draws from the predictions assuming a zero-inflated Poisson
distribution   (\cite{muchlinski_zero-inflated_2024}). Uncertainty estimates are not provided for the test set; hence they are produced by the organizers drawing from a Poisson distribution. 
\paragraph{\textbf{PaCE\_ShapeFinder	}}	A shape-based approach that uses Dynamic Time Warping that identifies historically analogous sequences of fatalities, takes the mean value of the data points that immediately follow each subsequence, and generates predictions by averaging the futures of historically similar sequences. The shape-based approach  stands out from traditional methods by analyzing entire time subsequences rather than isolated data points, thus enabling  a more comprehensive understanding of temporal dependencies in conflict data. Another advantage of the model is that it is able to visually represent complex temporal patterns, which  makes it particularly accessible for policymakers and analysts (\cite{schincariol_temporal_2024}).
\paragraph{\textbf{Brandt\_TW\_GLMM}} 
A Bayesian density forecast method using a Tweedie distribution and a Generalized Linear Model including country-month effects with auto-regression is fit over the training data. The model  is then updated for each forecast period to  include all prior training data. Alternative specifications of this model utilize a Negative Binomial (\textbf{Brandt\_NB\_GLMM}) or a Poisson distribution (\textbf{Brandt\_P\_GLMM}). Additional versions, building upon the above approach, include Generalized Additive Models (GAM) and country-month splines (\textbf{Brandt\_NB\_GAM}, \textbf{Brandt\_TW\_GAM}, \textbf{Brandt\_P\_GAM}) (\cite{brandt_bayesian_2024}).

\paragraph{\textbf{PLY\_cal\_double\_hurdle	}}	Three-stage hierarchical hurdle count model, where stages one and two predict whether any fatalities will occur at the country-month and pgm level, while stage three uses a truncated count regression model to predict the number of deaths conditional on the previous stages. By decomposing the prediction of fatalities into three respective problems that can be comprehended as two Bernoulli and one truncated Poisson random variables, an  estimate of the aleatoric uncertainty is obtained by  aggregating the stage-wise uncertainty. Platt scaling is utilized to enforce adequate representation of the target variable's relative frequency. This approach  offers full predictive distributions of future violence intensity at the pgm level and is easily interpretable, as it provides  substantive effect estimates for all violence predictors. Moreover, this  approach is computationally lightweight compared to most other machine learning algorithms \citep{fritz_predicting_2024}.
\paragraph{\textbf{Randahl\_Vegelius\_markov\_omm	}}	Observed Markov model (OMM) which utilizes
domain knowledge about conflict states to define observed states
through which countries can move over time. Uncertainty estimates are produced by propagating the state variable 15 steps into the future from the last data point in the training set, and randomly drawing fatalities at each time point conditional on the drawn state at that time point. Additional model specifications include a Hidden (pseudo-) Markov model (HPMM) (\textbf{Randahl\_Vegelius\_markov\_hpmm}), and a Gaussian process continuous Markov model (GPCMM) (\textbf{Randahl\_Vegelius\_markov\_gpcmm}). The first utilizes unknowable,
latent or hidden states through which countries can move over time, relaxing transition matrices assumptions, and the second utilizes a continuous observed Markov 'state' through which countries move over time (\cite{randahl_forecasting_2024}).
\paragraph{\textbf{UNITO\_transformer	}}	Pre-trained transformers incorporating an attention-based encoder, a temporal decoder and residual connections between input and output to preserve linear activation. The temporal decoder integrates information from the attention mechanism and directly from the past targets and future covariates through a residual connection, generating output incrementally for forecasting horizons. A Negative Log Likelihood (NLL) loss function is used to train the model, optimizing probabilistic forecasting by assuming a negative binomial distribution (\cite{macis_predicting_2024}).

Since the beginning of the challenge, the organizers made available a couple of simple benchmark models to provide some context for the evaluation metrics and a baseline for comparison during the model development phase. Following the recommendations by the scoring committee, we added a few benchmark models later on as a complement to the initial benchmarks. However, we do not use the latter benchmarks -- marked as ph (`post-hoc') -- as scoring rule as they were not disclosed to the participants from the start. The post-hoc benchmarks are still useful for informational purposes, as a simple heuristic that facilitates the comparison across models. Importantly, we note that the contribution of Becker \& Drauz (2024) has been based on the same logic of the \textit{conflictology} benchmarks since the opening of the prediction challenge,  before the organizers decided to add this benchmark. As such, we highlight that they were the first participants to apply this idea in this challenge and in our field. If this benchmark turns out to do well, their model is also likely to do well, and will be duly credited for their innovation.

\paragraph{\textbf{VIEWS\_bm\_exactly\_zero} \textbf{(cm/pgm})} This benchmark model forecasts zero fatalities for all country months or PRIO-GRID months.

\paragraph{\textbf{VIEWS\_bm\_last\_historical} (\textbf{cm/pgm})}  This benchmark model uses the last observed value (for 2018--2023, October for the preceding year, for 2024/25 April 2024) as the point prediction for a given country or grid cell, for each of the months in the forecasting window. A probability distribution is generated by drawing from the Poisson distribution with the point prediction as the mean and variance.

\paragraph{\textbf{VIEWS\_bm\_ph\_conflictology\_country12} (\textbf{cm)}} This benchmark model uses the set of values from each of the past 12 months for the same country as the forecast distribution for that country. Using empirical probability distributions as a benchmark has long been a common practice in meteorology (\cite{Murphy1984JotASA}).

\paragraph{\textbf{VIEWS\_bm\_ph\_conflictology\_neighbors12} (pgm)} This benchmark model uses the set of values from each of the past 12 months for the same grid cell and the immediate neighbors of that grid cell as the forecast distribution for that grid cell. 

\paragraph{\textbf{VIEWS\_bm\_conflictology\_bootstrap240} (\textbf{cm)}} A benchmark conflictology model of bootstrapped samples, randomly drawing a set of values from the past 240 months (20 years) from any country in the world as the forecast for a given country. The forecasted distribution is consequently similar for all countries. The model should be well calibrated at a global level but perform very poorly for individual country months.

\section{Evaluation and metrics}
\subsection{Scoring committee}
The evaluation of the contributions will be done by a scoring committee consisting of 3--4 members from the forecasting expert community as well one or two representatives from the user community. The scoring committee will provide an independent evaluation of the models, based primarily on the quantitative scoring outlined below (which, for technical reasons, will be computed by the ViEWS team), but also on the short summaries of the contributions, as well as any additional criteria the committee itself may deem relevant. The role of the scoring committee is thus to provide a comprehensive assessment of the contributions that goes beyond the quantitative evaluation, and account for additional aspects of the models -- such as innovation, or replicability -- that may represent a valuable contribution despite not being necessarily rewarded by the quantitative scoring.  We will place approximately equal weight on the joint performance across the test period predictions and the predictions for 2024, the true future.

The scoring committee will consist of Philip Schrodt (Parus Analytics, former Pennsylvania State University), Céline Cunen (Norwegian Computing Center and University of Oslo), Thomas Mayer (Preview, German Ministry of Foreign Affairs), and Seth Caldwell (United Nations Office for the Coordination of Humanitarian Affairs -- UN OCHA).

\subsection{Evaluation guiding principles}
As argued by Tillman Gneiting, `[s]ingle-valued point forecasts... can lead to grossly misguided inferences, unless the scoring function and the forecasting task are carefully matched' \citep[][p. 746]{Gneiting2011JASA}.\footnote{\citet{Gneiting2011JASA} further argues that `[e]ffective point forecasting requires that the scoring function be specified a priori, or that the forecaster receives a directive in the form of a statistical functional', and that `it is critical that the scoring function [is] consistent for [the functional], in the sense that the expected score is minimized when following the directive' (Ibid.). What this means is that if we want to use single-valued point forecasts, we have to a priori decide on an elicitable target to summarise  the predictive distribution, e.g., the mean. We can then find a scoring function that is consistent with this target. In the case of the mean, that would be the square error.} 
As the mean is not representative of the entirety of a distribution, especially when it is zero-inflated and skewed as in the case of interest, 
commonly used metrics such as the Mean Squared Errors which heavily rely on the mean should be discarded in favour of broader evaluation criteria. These criteria should account not only for a certain target representation of the predictive distribution (such as the mean), but fully encompass a general characterization of how well the predictive distribution fits observed outcomes.\footnote{We will define more clearly what we mean by \emph{fit} below.}

For the challenge to be useful for a wider range of researchers, policy-makers, and NGOs, we therefore need to calculate different evaluation metrics for distinct use cases, accounting for different aspects of the predictive distributions, and depending on various goals. A main goal of probabilistic forecasts evaluation, for example, is to reward the `sharpness of the distribution, subject to calibration' \citep[][p. 359]{Gneiting:2007JASA}. Other evaluation metrics may reward different aspects of the forecasts. If we are only interested in `one-shot forecasts' -- how well the model does at predicting specific single events -- we would mainly evaluate predictions at the part of the predictive distribution that is deemed the most likely outcome. If we are interested in performance over many tries, then calibration becomes more relevant. If we are only interested in extreme outcomes, then predictive performance at the tails of the predictive distribution becomes more relevant. Additionally, if the costs and benefits of prediction depend on the location of events in space and time relative to the predicted distribution, then we would want to calculate relative performance of models in that spatial-temporal context. 
In addition to sharpness and calibration, other criteria need to be specified to define useful predictive distributions in conflict fatalities: \textit{focus}, \textit{nearness}, and \textit{propriety}. We discuss these criteria and the corresponding evaluation metrics below.

\subsection{What to reward: Evaluation criteria and metrics}
Here, we present the main metrics used to evaluate the contributions, our motivations for their relative importance, and discuss some practicalities of evaluation and  power/data sparseness concerns.

We will use the following notation in what follows:
\begin{itemize}
    \item $f(x)$: The forecast distribution/mass function
    \item $F(x)$: The forecast cumulative distribution function
    \item $y$: The observed value
\end{itemize}
As noted above, we will do the evaluation in terms of $y$ as the non-logged count of fatalities, in place of the  $ln(y+1)$ used in \citet{Hegre2022II}. As shown in Figure \ref{fig:FatalityDistribution}, the distribution of the target variable is challenging, being zero-inflated and heavy-tailed. The evaluation metrics we propose below should be well equipped to evaluate predictive distributions seeking to match this target.

Note that we will evaluate the predictions for 2024 against the VIEWS aggregations of the UCDP-Candidate data \citep{Hegre2020RP}, since these are the only data that will be available at that time. For the other test periods, we will evaluate against the final UCDP-GED data \citep{Sundberg2013JPR} aggregated at the two VIEWS levels of analysis.

The metrics are designed to reward five qualities of probabilistic forecasting systems -- calibration, sharpness, focus, nearness, and propriety. 

Table \ref{tab:Qualities} summarizes these five and link them to the main metrics that we will be using.
\begin{table} 
    \centering
    \begin{tabular}{|l|l|l|l|l|l|}
     \hline
     & \multicolumn{5}{c|}{Desirable qualities} \\ \hline
        Rule & Calibration & Sharpness & Focus & Nearness & Propriety \\ \hline
        CRPS &    X & X & - & x & X \\
        IGN  &    - & X & X & - & x \\
        MIS &    X & X & - & - & X \\ \hline
    \end{tabular}
    \caption{Beneficial qualities of probabilistic forecasting systems and how they are assessed by core evaluation metrics. Large \textit{X}-s mean the metric is highly useful for assessing the respective quality. Small \textit{x}-s mean the metric captures the quality partially or with conditions.\protect\footnote{CRPS gets a little $x$ for distance-sensitivity because it is sensitive to misses within observation bins (across values for an observation), but not across observations.}}
    \label{tab:Qualities}
\end{table}
\paragraph{Calibration} A model is well calibrated when the predicted frequency of $y$-values corresponds to the observed frequency of $y$ in new data (this is a joint function of $f(x)$ and $y$). For instance, if a well-calibrated model predicts a 30\% probability of at least 100 deaths for 10 observations, in the actual data 3 out of the 10 observations recorded at least 100 deaths.

\paragraph{Sharpness} Concentrated predictive distributions are preferred as they encode more information, as defined by \citet{Shannon1948}, across all possible values of $y$. 
A model that predicts with 80\% probability that the true value is between 20 and 50 is preferred to one that specifies the 80\% prediction interval to be between 10 and 100, everything else equal. \citet{Gneiting2007JRSS} argue that the ideal forecast should maximize sharpness subject to calibration.

\paragraph{Focus} refers to the aim that useful predictive distributions should provide high plausibility at the exact value (or as close to the exact value as possible) that materialize. This quality is a function of $f(x)$ evaluated at $y$. Focus is sometimes referred to as locality. To see the value of focus, as distinct from calibration and sharpness, imagine the forecast distribution as a flashlight pointed at an infinite ruler that runs from 0 to infinity (or some very large number if you prefer). As the plausibility of specific values for the observation (from the point of the view of the forecast) increase, more and more light is thrown on those values on the ruler. Where calibration would ask if the intensity of light across the ruler matches what we would see across many observations and sharpness would simply measure intensity of light irregardless of the values on the ruler, focus suggests we look at high much light is focused exactly on the observed value. Thus, we are not interested in the light away from the actual value. Indeed without valuing focus/locality one can priviledge models that assign much less plausibility to actual values solely because they are better calibrated in areas far away from the realized outcome. Focus corresponds to the argument that evaluation of the full predictive distribution is not warranted, as `when assessing the worthiness of a scientist's final conclusions, only the probability he attaches to a small interval containing the true value should be taken into account' \citep[Bernardo 1979 p. 689, cited in][365f.]{Gneiting:2007JASA}.

\paragraph{Nearness} Predictive distributions should array the plausibility of future values near the actual values, based on a usefully applied geometry. In our context, `near' means close in time \textit{and} space -- predicting violence three months too early is better than predicting twelve months too early, and predicting violence 100 km from where it happens is better than 1,000 km off. This quality is called `sensitivity to distance' by \citet{Gneiting:2007JASA}.

\paragraph{Propriety} encourages the reporting of predictive distributions that represent the honest beliefs of the forecaster or model. Proper scoring rules accomplish this by ensuring that the maximization of the expected reward for the forecaster occurs when reporting their underlying beliefs and not bending the shape of those beliefs in a particular direction \citep{Gneiting2007JRSS, Czado2009B}. In contrast, an improper score might reward increased certainty or a shifted mode for the distribution to hedge relative to the true underlying beliefs.\footnote{A survey of the use of proper scoring functions across different scientific domains can be found in \citep{Carvalho2016},
and a discussion with specific reference to count data is contained in \citet{Czado2009B}.} While propriety might seem like it should always and everywhere apply, as with each of the other traits, there are trade-offs. For example, in the current domain of the competition directly measuring focus with a proper scoring rule like the the raw log score (defined on the full count sample space and not just a coarser range) is not possible without computing infinite penalties regularly.


\subsection{Metrics}
The scoring committee will consider the metrics below when evaluating the contributions. These metrics reward different properties of the forecasting distributions, as outlined in the previous section. The main scoring and ranking of the contributions will be done in terms of the CRPS. The other metrics will be used for secondary scoring and rankings, and facilitate a richer discussion of model performance. The code implementing the evaluation, including all the detailed adaptations reviewed below, is found in \url{https://github.com/prio-data/prediction_competition_2023} .

\subsubsection{Main metric: Continuous Rank Probability Score (CRPS) }
\begin{figure}[htp]
    \centering
\includegraphics[width=.6\linewidth]{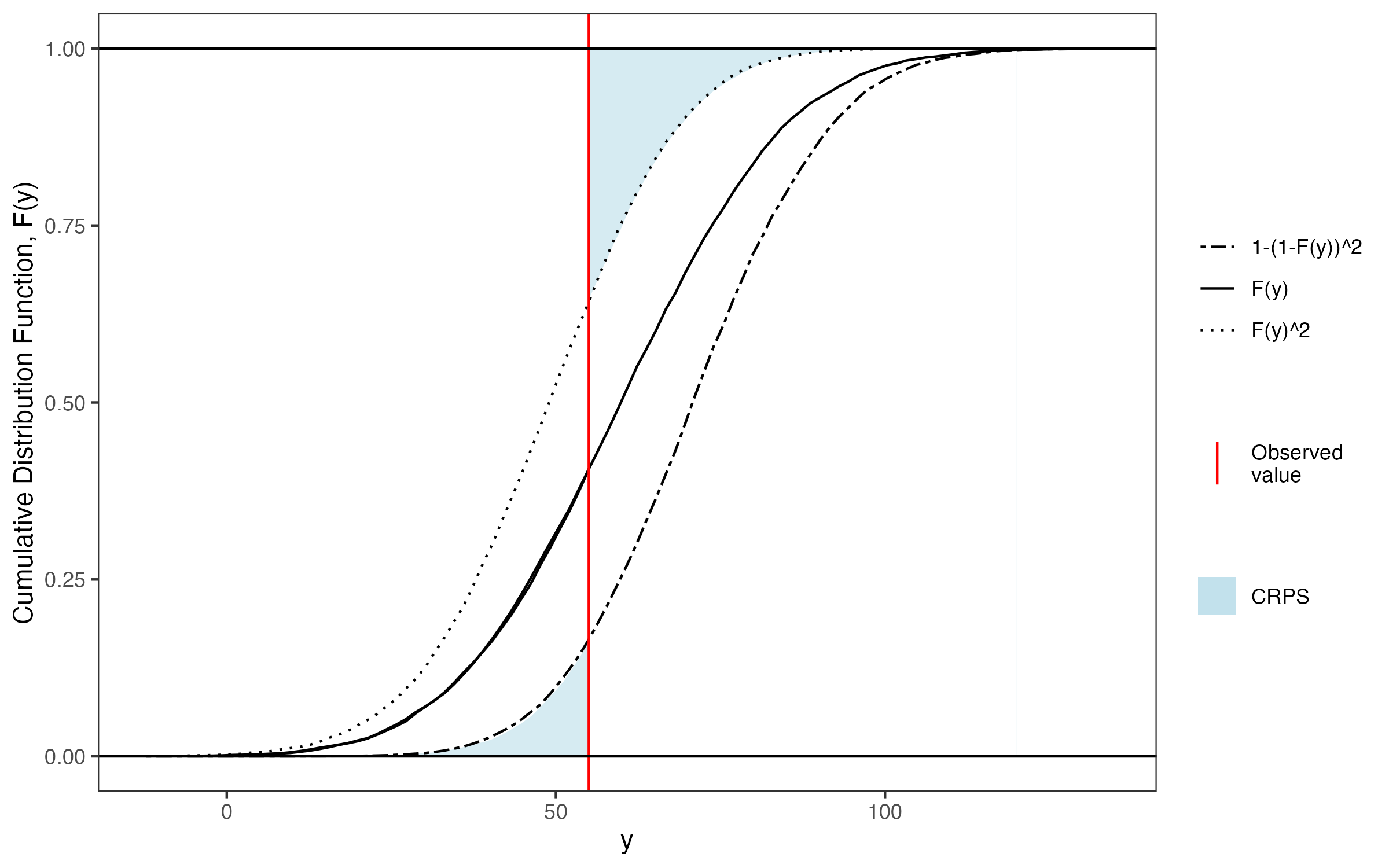}
    \caption{Illustration of the contribution to the CRPS metric for one instance from the observed data, based on \citet{Bracher2021PLOS}}
    \label{fig:crps}
\end{figure}
 CRPS values sharpness subject to calibration, and is an assessment of the full forecast distribution given the outcome. The continuous rank probability score for an individual observation is defined as:
 \[ CRPS(F_{i},y_{i}) = \int_{\mathbb{R}} (F_{i}(x)-\mathbbm{1}(x - y_{i}))^2 dx \]
where $\mathbbm{1}(z)$ is the indicator function (or Heaviside step function), defined as
\begin{equation}
  \mathbbm{1}(z) = 
    \begin{cases}
      1 & \text{if $z\ge 0$} \\
      0 & \text{otherwise}
    \end{cases}       
\end{equation}
Figure \ref{fig:crps} illustrates the metric. $y_{i}$ is the observed value, the red line the indicator function $\mathbbm{1}(z)$, and the black line the cumulative probability distributions. The blue shade is the contribution to the metric for this prediction. CRPS will be minimized when the forecast distribution has low variance, but only if it is centered around the actual values. Therefore, averaged over test-set observations, we calculate the mean CRPS for an entry to reward calibration (ie does the shape of the CDF match the shape of the actual values) and sharpness (ie CDFs that approach the true step-function are preferred). In addition, the CRPS is proper.    

One can think of the Continuous Rank Probability Score as a generalization of the Brier score to infinitely small bins.  Broadly, CRPS is a generalization of the MAE for any predictive distribution: if CRPS is used to compare a `point' prediction as a CDF with a point observation, it gives MAE.\footnote{For finite count data, the CRPS can also be considered a Ranked Probability score \citep{Czado2009B, Kolassa2016IJF}. We will follow the standard convention and refer to the metric as CRPS here as the integral does apply to the CDF of the counts as described above. In addition, the two representations will lead to different normalizations that will not change ranks of models but will alter the absolute value of the scores computed.} Unlike the Ignorance Score (see below), CRPS takes the distance between the forecasted plausibility of values and the observed value into account across the whole range of values. CRPS like many other performance metrics is measured on a scale that depends on the variance of the target observations. This means that the computed values cannot be compared across different test-set samples without extremely strong and likely misleading assumptions.

\paragraph{Implementation:}
We compute the measure using the \texttt{properscoring.crps\_ensemble()} function/package as implemented in \texttt{xskillscore.crps\_ensemble()}. We will weigh each sample forecast equally. This approach uses the Empirical CDF to elicit probabilities (see \citet{Kruger2021ISR}.

\subsubsection{Secondary metric I: Log score/ignorance score}
The log score (also called ignorance score) is the log of the predictive density evaluated at the actual observation:
 \[ IGN(f,y)=-log_2(f(y)) \]
\citet{Winkler1969JASA} showed that in the case where a forecaster is only rewarded for the probability assessment of the actualized outcome and there are more than two outcomes, the only proper scoring function (i.e., the only scoring function that does not encourage "dishonest" elicitations) is the log score. Whether it is important to keep the forecasters honest, is another question. Since we are also eliciting the full forecast distribution with other proper scores, if we chose to use another scoring function for the local assessment, the forecaster would need to weigh between optimizing payout for the full forecast distribution versus the local forecast. Chances are high that none in our competition are thinking about the forecast in this game-theoretic sense in any case. Other scoring functions can be used to evaluate local forecasts (e.g., the linear score or the quadratic score). These could be simpler to calculate, e.g., if someone assign 0 probability to an outcome.

The ignorance score complements the CRPS by scoring the predicted probability of the observed event, instead of the distance between the predicted and observed. Therefore it emphasizes how much belief is focused at the observed value (see implementation details below however).

\paragraph{Implementation:}

It not always straight-forward to elicit a predicted probability for any outcome. Since we base our computations on samples and not full probability distributions, we need some way to translate samples into probabilities. There are many ways to do this, however. \citet{Kruger2021ISR} show that kernel density estimation is an unstable approach for the Log Score, and this is particularly true for zero-inflated data such as ours. Our preferred way is to bin the predicted and observed outcomes, upsample the  and then add each of the possible binned outcomes to each forecast sample. This way, we will guarantee being able to elicit non-zero probabilities for each outcome. This approach is similar to assuming a naive uniform prior distribution. For the addition of each binned outcomes to be fair, the number of samples must be the same for all. We up- (or down-)sample the forecasts to 1000 samples before binning and adding each possible outcome once.

We will use the following binning scheme: 
0, 1--2, 3--5, 6--10, 11--25, 26--50, 51--100, 101--250, 251--500, 501--1000, 1001--.

E.g., if the forecast is [0,0,0,2,11,4], this gets up-sampled to 1000 samples, then binned according to the above scheme. Finally, the 11 possible outcomes are added, yielding 1011 samples. If the observed value is 10, the log score is 2.46, if it is 0, the log score is 0.92. By having a maximum number of samples, we also limit how certain any forecast can be. E.g., with 11 bins, we can at most be right 1001 out of 1011 times (0.99\%, or Ignorance score of 0.014).

A problem with binning is that it no longer becomes a proper score. Indeed, we could have used the probability directly here as our score (i.e., linear), since there are other ways to game this score (e.g., shifting samples slightly in the direction of the bins with the mode). With the same up-sampling procedure, if true is 0, [0,0,0,2,11,4] yields 0.53 (slightly above the exact 0.5), and if true is 10, the forecast yields 0.18 (slightly above the exact 0.1667). The deviance is created from up-sampling (using scipy.signal.resample) and adding the bins as forecasts.

\subsubsection{Secondary metric II: Mean Interval Scores (MIS)}
The M4 competition \citep{Makridakis2018IJF} use the Mean Scaled Interval Scores (MSIS). MSIS is set up as a battle between making the prediction interval (based on a lower and upper quantile) as small as possible whilst still ensuring a good coverage rate. It does not consider the mass of the predictive distribution within the interval, so it is not an accuracy metric like CRPS. Still, it performs quite similarly to CRPS, and particularly for forecast samples that are provided from approaches that predict quantiles. 
Unlike the CRPS, MSIS focuses on the most likely values, without narrowing to a point, and penalizes large prediction intervals while rewarding coverage. The scaling in MSIS is used to make the measure scale-independent as the M4 competition deals with a large set of different types of time-series with varying time-scales and variability. Since this is not needed here, we have simplified this score to just the Mean Interval Score (MIS), which is also discussed in \citep{Gneiting:2007JASA}. The Interval Score is defined as:
\[ IS_{it} = (U_{it} - L_{it}) + \frac{2}{a}(L_{it} - Y_{it})\mathbbm{1}(L_{it}-Y_{it}) + \frac{2}{a}(Y_{it} - U_{it})\mathbbm{1}(Y_{it} - U_{it}) \]
where $U_{it}$ and $L_{it}$ are the upper and lower prediction sample quantiles using the set prediction interval, $a = [1 - ($prediction interval$)]$ (e.g. for a 95\% prediction interval, $a = 0.05$) and $\mathbbm{1}(z)$ is the indicator function as defined previously. To get the Mean Interval Score, the $IS_{it}$ is averaged across time $t$ and units $i$. The three terms could be called "interval width", "lower coverage", and "upper coverage". If the outcome is above the interval, then the upper coverage is high, and vice versa if the outcome is below the interval. If the outcome is within the interval, the score is the interval width (a perfect prediction would still yield 0 score).

For example, if your prediction is [0, 0, 4, 10] and the outcome is 5, if $a=0.1$ then L is 0 and U is 8.2, and since the outcome is within the interval, IS is 8.2 - 0  = 8.2.

\paragraph{Implementation:}

It is not straight-forward to calculate quantiles from a sample \citep{Hyndman1996AS}. We use the linear (Gumbel) method for interpolation, which is the default approach in NumPy. We estimate the MIS for the 90\% prediction interval (a = 0.1). We consider this mainly a calibration metric, where we want to measure the ability to issue forecasts that most of the time includes the observed outcome, but it also contains information about sharpness (although samples just within the forecast interval are penalized just as little as those exactly at the outcome). Since CRPS is already accounting for the error-distance aspect, we opted for as wide prediction intervals as we deemed possible to elicit. Through simulation, we found that the ability of forecasts with 1000 samples to accurately provide estimates of quantiles for overdispersed distributions outside the 90\% prediction interval tapers off quickly, which is why we did not settle for a wider interval. It is also at the tail end of the distribution that the choice of the quantile interpolation method matters the most. By setting this to 90\%, we are attempting to reduce the effect of the computation and number of samples -- which would have a potentially large effect on wider intervals -- to learn about the range of the most likely values that extend towards the tails of the count distribution.

\section{Benchmark model evaluation}
Table \ref{tab:cm_benchmarks} shows evaluation scores for the four benchmark models described above, for each of the years 2018--2023 for which we have historical data, for each of the three metrics under consideration. We also show the average scores across the six years in the row labeled `overall'. Table \ref{tab:cm_bm_exactly_zero} shows the scores for the benchmark model \textbf{VIEWS\_bm\_exactly\_zero} where all units are forecasted as zero fatalities. Mean scores across all six years are 56.84 for CRPS, 1.59 for IGN, and 1136.80 for MIS. Reflecting the steady escalation of violence levels since 2018 \citep{Davies2023JPR}, the model predicting no violence anywhere does worse for the latest years -- obviously, an exactly-zero model would not have been able to forecast the escalation of violence in Ukraine, Ethiopia, and Sudan in 2021--22. Table \ref{tab:cm_bm_last_historical} shows the scores for the \textbf{bm\_last\_historical} model that predicts the violence observed in the last month with data will continue unchanged (with some added uncertainty). For the first three years, this model does better than the exactly zero model, but for 2021--23 it is even more surprised by the new wars than the zero model.
\begin{table}[htp]
    \centering
    \begin{subtable}[b]{0.49\textwidth}
    \begin{tabular}{lrrr}
         & crps  & ign  & mis \\ \hline
    2018 & 24.13 & 1.56 & 482.61 \\
    2019 & 23.02 & 1.56 & 460.38 \\
    2020 & 32.04 & 1.55 & 640.81 \\
    2021 & 87.34 & 1.61 & 1746.78 \\
    2022 & 120.97 & 1.63 & 2419.36 \\
    2023 & 53.54 & 1.61 & 1070.86 \\ \hline
    Overall & 56.84 & 1.59 & 1136.80 \\ \hline
    \end{tabular}
    \caption{bm\_exactly\_zero} \label{tab:cm_bm_exactly_zero}
    \end{subtable} 
    \begin{subtable}[b]{0.49\textwidth}
    \begin{tabular}{lrrr}
         & crps  & ign  & mis \\ \hline
    2018 & 20.17 & 1.20 & 380.62 \\
    2019 &  9.48 & 1.05 & 172.69 \\
    2020 & 23.70 & 1.11 & 455.81 \\
    2021 & 85.61 & 1.23 & 1690.71 \\
    2022 & 131.02 & 1.12 & 2599.28 \\
    2023 & 678.96 & 1.12 & 13523.46 \\ \hline
    Overall & 158.16 & 1.14 & 3137.09 \\ \hline
    \end{tabular}
    \caption{bm\_last\_historical} \label{tab:cm_bm_last_historical}
    \end{subtable} \\
\vspace{5mm}
    \begin{subtable}[b]{0.49\textwidth}
    \begin{tabular}{lrrr}
         & crps  & ign  & mis \\   \hline
    2018 & 14.48 & 0.64 & 186.55 \\
    2019 &  9.15 & 0.61 & 89.06 \\
    2020 & 21.34 & 0.57 & 344.96 \\
    2021 & 76.85 & 0.69 & 1435.55 \\
    2022 & 124.00 & 0.69 & 2142.13 \\
    2023 & 50.36 & 0.68 & 1042.92 \\ \hline
    Overall & 49.36 & 0.65 & 873.53 \\ \hline
    \end{tabular}
    \caption{VIEWS\_bm\_ph\_conflictology\_country12} \label{tab:cm_bm_conflictology_country12}
    \end{subtable} 
    \begin{subtable}[b]{0.49\textwidth}
    \begin{tabular}{lrrr}
         & crps  & ign  & mis \\ \hline
    2018 & 23.58 & 1.12 & 454.09 \\
    2019 & 22.46 & 1.11 & 426.01 \\
    2020 & 31.42 & 1.12 & 606.00 \\
    2021 & 86.63 & 1.15 & 1708.30 \\
    2022 & 120.25 & 1.15 & 2380.74 \\
    2023 & 52.72 & 1.15 & 1030.99 \\ \hline
    Overall & 56.17 & 1.14 & 1101.02 \\ \hline
    \end{tabular}
    \caption{VIEWS\_bm\_conflictology\_bootstrap240} \label{tab:cm_bm_conflictology_global240}
    \end{subtable} 
    \caption{Benchmark model evaluation, \textit{cm} level, four benchmark models, 2018--2023.}
    \label{tab:cm_benchmarks}
\end{table}
\textbf{VIEWS\_bm\_ph\_conflictology\_country12} (Table \ref{tab:cm_bm_conflictology_country12}) is the strongest of the benchmarks -- using the historical observations for the last 12 months as the prediction distribution, it performs much better on all metrics. The scores equal to 49.36 for CRPS, 0.65 for IGN, and 873.53 for MIS.

Table \ref{tab:cm_bm_conflictology_global240} scores the final \textit{cm} benchmark, the \textbf{VIEWS\_bm\_conflictology\_bootstrap240} model that predicts that all country months have the same probability distribution as the global record back to the late 1990s. Being just as uncertain and pessimistic as the exactly zero benchmark is confidently optimistic, it does considerably better. In terms of CRPS, it is actually the best model for 2022.
\begin{table}[htp]
    \centering
    
    \begin{subtable}[b]{0.32\textwidth}
    \begin{scriptsize}
        
    \begin{tabular}{lrrr}
         & crps  & ign  & mis \\ \hline
    2018 &	0.14	&	0.09	&	2.89	\\
2019 &	0.12	&	0.09	&	2.31	\\
2020 &	0.13	&	0.11	&	2.64	\\
2021 &	0.94	&	0.12	&	18.80	\\
2022 &	1.14	&	0.12	&	22.75	\\
2023 &	0.22	&	0.12	&	4.47	\\ \hline
    Overall & 	0.45	&	0.11		& 8.98	\\ \hline
    \end{tabular}
    \end{scriptsize}
    \caption{\scriptsize{VIEWS\_bm\_exactly\_zero}} \label{tab:pgm_bm_exactly_zero}
    \end{subtable} 
    \begin{subtable}[b]{0.32\textwidth}
    \begin{scriptsize}
    \begin{tabular}{lrrr}
         & crps  & ign  & mis \\ \hline
    2018 &	0.39	&	0.12	&	7.15	\\
2019 &	0.14	&	0.11	&	2.62	\\
2020 &	0.16	&	0.12	&	2.99	\\
2021 &	0.97	&	0.13	&	19.08	\\
2022 &	1.46	&	0.15	&	28.53	\\
2023 &	9.75	&	0.15	&	193.97	\\  \hline
    Overall & 2.15	&	0.13	&	42.39	\\  \hline
    \end{tabular}
    \end{scriptsize}
    \caption{\scriptsize{VIEWS\_bm\_last\_historical}}\label{tab:pgm_bm_last_historical}
    \end{subtable} 
    \begin{subtable}[b]{0.32\textwidth}
    \begin{scriptsize}
    \begin{tabular}{lrrr}
         & crps  & ign  & mis \\ \hline
   2018 &	0.19	&	0.08	&	2.83	\\
2019 &	0.12	&	0.08	&	1.89	\\
2020 &	0.13	&	0.08	&	2.07	\\
2021 &	0.93	&	0.09	&	17.87	\\
2022 &	1.14	&	0.10	&	22.28	\\
2023 &	0.52	&	0.10	&	13.22	\\ \hline
    Overall & 0.51	&	0.09	&	10.03	 \\ \hline
    \end{tabular}
    \end{scriptsize}
    \caption{\scriptsize{VIEWS\_bm\_ph\_conflictology\_country12}} \label{tab:pgm_bm_conflictology_country12}
    \end{subtable} 
    \vspace{5mm} \\
    
    \begin{subtable}[b]{0.32\textwidth}
    \begin{scriptsize}
    \begin{tabular}{lrrr}
         & crps  & ign  & mis \\  \hline
   2018 &	0.15	&	0.08	&	3.06	\\
2019 &	0.11	&	0.08	&	1.88	\\
2020 &	0.12	&	0.08	&	2.12	\\
2021 &	0.93	&	0.10	&	18.11	\\
2022 &	1.13	&	0.10	&	22.48	\\
2023 &	0.25	&	0.10	&	4.03	\\ \hline
    Overall & 0.45	&	0.09	&	8.61	\\ \hline
    \end{tabular}
    \end{scriptsize}
    \caption{\scriptsize{VIEWS\_bm\_ph\_conflictology\_neighbors12}}\label{tab:pgm_bm_conflictology_neigh12}
    \end{subtable} 
    \begin{subtable}[b]{0.32\textwidth}
    \begin{scriptsize}
    \begin{tabular}{lrrr}
         & crps  & ign  & mis \\ \hline
    2018 &	0.14	&	0.09	&	2.89	\\
2019 &	0.12	&	0.10	&	2.31	\\
2020 &	0.13	&	0.11	&	2.64	\\
2021 &	0.94	&	0.12	&	18.80	\\
2022 &	1.14	&	0.12	&	22.75	\\
2023 &	0.22	&	0.12	&	4.47	\\  \hline
    Overall & 0.45	&	0.11	&	8.98 \\ \hline
    \end{tabular}
    \end{scriptsize}
    \caption{\scriptsize{VIEWS\_bm\_conflictology\_bootstrap240}} \label{tab:pgm_bm_conflictology_global240}
    \end{subtable} \\
    \caption{Benchmark model evaluation, \textit{pgm} level, five benchmark models, 2018--2023.}
    \label{tab:pgm_benchmarks}
\end{table}
Table \ref{tab:pgm_benchmarks} scores the five benchmark models at the \textit{pgm} level. Four of them are constructed along the same lines as the \textit{cm} benchmarks. In addition, we have included a model \textbf{VIEWS\_bm\_ph\_conflictology\_neighbors12} that takes the observed values in the cell as well as its first order neighbors as the `draw' from the prediction distribution. 

The \textbf{VIEWS\_bm\_exactly\_zero} model (Table \ref{tab:pgm_bm_exactly_zero}) has average scores of 0.45 for CRPS, 0.11 for IGN, and 8.98 for MIS. Just as at the country level, the scores are much worse for the years 2021--22 than for the first three years. The extreme zero-inflation of the fatality count \textit{pgm} level makes the exactly-zero model hard to beat. The \textbf{VIEWS\_bm\_last\_historical} obtains worse scores across all metrics. The \textbf{VIEWS\_bm\_ph\_conflictology\_country12} also performs worse than the optimistic \textbf{VIEWS\_bm\_exactly\_zero} model for both CRPS and MIS and slightly better for IGN. The variant of the conflictology benchmark model that also includes the observations for the immediate neighbors of each grid cell, on the other hand, does considerably better than the exactly zero model. The fundamentally uncertain \textbf{VIEWS\_bm\_conflictology\_bootstrap240}
model performs very similarly to the exactly zero model -- since 99.5\% of the historical values are zero, the predictions are almost identical.

\subsection{Ensembling}
We will generate an ensemble from all contributed models, weighted by some function of the CRPS for the test set. Building on the ensemble, we can assess models based on their unique or distinct contribution relative to the full set of all models, or their diversity relative to the average contribution, as well as evaluate the joint contribution of all models.

\section{Conclusions}
This article draft has presented the structure of the VIEWS 2023/24 prediction challenge. The forecasts for the true future will be posted at \url{https://predcomp.viewsforecasting.org} at the end of June 2024, and be subject to evaluation when data on actual conflicts become available. As such, this draft manuscript has a function analogous to a pre-analysis plan: a statement of the forecasting models made publicly available before the true future prediction window commences. The evaluation of the models will be reported in a second article from the challenge, to be finalized in the Fall of 2025, after the end of the true future forecasting window.

\section{Acknowledgements}
The VIEWS Prediction Challenge 2023/2024 received financial support by the German Ministry for Foreign Affairs. We acknowledge the support of the PREVIEW Team and are particularly grateful to Ida Bauer and Thomas Mayer for their help in organizing the workshop in Berlin.

\bibliographystyle{unsrtnat}
\bibliography{main}  
\end{document}